\documentclass[english,times,pra,twocolumn,superscriptaddress]{revtex4-1}
\usepackage{amsmath}
\usepackage{amssymb}
\usepackage{graphicx}
\usepackage{times}

\begin{document}
\title{Ramsey-biased spectroscopy of superconducting qubits under dispersion}
\author{Yan Zhang}
\affiliation{Institute of Applied Physics and Materials Engineering, University
of Macau, Macau, China}
\author{Tiantian Huan}
\affiliation{Institute of Applied Physics and Materials Engineering, University
of Macau, Macau, China}
\author{Ri-gui Zhou}
\affiliation{{\small{}College of Information Engineering, Shanghai Maritime University,
Shanghai 201306, China}}
\author{Hou Ian}
\email{houian@um.edu.mo}

\affiliation{Institute of Applied Physics and Materials Engineering, University
of Macau, Macau, China}
\begin{abstract}
We proposed a spectroscopic method that extends Ramsey's atomic spectroscopy
to detect the transition frequency of a qubit fabricated on a superconducting
circuit. The method uses a multi-interval train of qubit biases to
implement an alternate resonant and dispersive couplings to an incident
probe field. The consequent absorption spectrum of the qubit has a
narrower linewidth at its transition frequency than that obtained
from constantly biasing the qubit to resonance while the middle dispersive
evolution incurs only a negligible shift in detected frequency. Modeling
on transmon qubits, we find that the linewidth reduction reaches 23\%
and Ramsey fringes are simultaneously suppressed at extreme duration
ratio of dispersion over resonance for a double-resonance scheme.
If the scheme is augmented by an extra resonance segment, a further
37\% reduction can be achieved.
\end{abstract}
\maketitle

\section{Introduction}

Josephson-junction devices fabricated with superconducting materials
have been used extensively to implement the fundamental quantum data
unit -- qubit -- on a solid-state platform, giving rise to superconducting
qubits suitable for quantum computing applications~\cite{Mooij1999,Clarke2008}.
Though having different structural types (major ones being charge
qubit~\cite{nakamura97}, flux qubit~\cite{Orlando1999,vanderWal00},
and phase qubit~\cite{martinis02}), a superconducting qubit is basically
an anharmonic multi-level system, which under a low-temperature environment
is approximately two-level. When an incident microwave field of frequency
that matches the transition frequency between the two levels, the
qubit is excited and undergoes Rabi oscillations. Therefore, being
the main characteristics of the solid-state system, this transition
frequency is comparable to those between atomic orbital levels in
natural atoms, making the superconducting device an artificial atom~\cite{Kastner1993,you05}.

Transition frequencies of superconducting qubits usually ranges between
$3$~GHz and $10$~GHz~\cite{Chiorescu2003,Wallraff2004}, which
fall into the milli-meter wave band and are typically determined by
feeding a continuous-wave (CW) microwave field generated from a vector
network analyzer (VNA) and measuring the transmission coefficient.
The CW field acting as a probe is fed into a waveguide and made resonant
with the qubit on the superconducting circuit~\cite{shen05}. The
output is connected to the receiving port of the VNA, resulting in
a sweep plot showing the absorption spectrum of Lorentzian distribution~\cite{Hime2006,Shevchenko2008},
similar to those of natural atoms.

The Lorentzian distribution is derivable from the standard Rabi oscillation
model, which is experimentally determined by sending a focused beam
of atoms through a cavity field that oscillates in a fixed resonance
zone~\cite{Torrey1941}. The resonance linewidth of the Lorentzian
is lower bounded by the coupling strength of the cavity field to the
atoms. By separating the resonance zone into two distant ones between
which the atoms are allowed to evolve freely, Ramsey discovered in
1950 that the linewidth of alkaline atoms adopts a non-Lorentzian
shape with a narrower full width at half maximum (FWHM)~\cite{Ramsey1950}.
The linewidth reduction can reach as high as 40\% even though two
side fringes are unavoidably added.

Ramsey's spectroscopy method is recently applied to the spectroscopy
of phase shifts of a micromirror~\cite{Abele2010} and further refined
by introducing more individually controlled resonance zones~\cite{Zanon2015}.
In this paper, we generalize Ramsey's method of separating fields
to superconducting circuit systems by inverting the roles of atoms
and resonating fields. In other words, the atoms become the qubit
fixated in the circuit whereas the standing cavity fields become a
microwave pulse traveling in the circuit waveguide. Again, the spectroscopy
of the qubit can be derived from the transmission coefficient between
the input and the output pulse amplitudes.

To be exact, we model after a transmon qubit~\cite{Koch2007}, which
is a flux-biased variant of charge qubit and less sensitive to charge
noise, for its wide applicability to quantum information and computation~\cite{Filipp2009}.
The Hamiltonian of the qubit, a two-junctioned superconducting loop,
is contributed by the supercurrent through the two parallel junctions
and the charge stored in the two equivalent capacitors. Hence, the
transition frequency depends partially on the magnetic flux through
the loop, which is externally tuned by a neighbouring current-controlled
inductor. To implement Ramsey's double-resonance scheme, we consider
tuning the qubit through the current-bias line while feeding the waveguide
input end with a microwave square pulse. During the qubit-pulse interaction,
the qubit is to operate alternatively in both the resonant and the
dispersive regimes by close-detuning and far-detuning the qubit. The
spectroscopy reflected in the transmission is then optimized by fine
tuning the respective durations of the two distinct regimes during
the interaction. We employ stochastic optimization that follows Maxwell
distribution~\cite{Walstad2013} in the total pulse duration to further
optimize the linewidth of the absorption spectrum. We note that the
method of double-segment resonances have been widely used in determining
the dephasing times of superconducting qubits~\cite{Chiorescu2003,fyan12,fyan13}.
Named Ramsey interferometry, it uses fixed $\pi/2$ resonance durations
to bring the qubit to and away from the $XY$-plane of a Bloch sphere.
In constrast, we vary the resonance durations here to minimize the
off-resonance transitions.

In the following, we develop in Sec.~\ref{sec:model} the formulas
for qubit evolution, from which the transition probability under a
double-resonance detection scheme is computed in Sec.~\ref{sec:double_res}.
We find a 23\% reduction in FWHM over conventional CW detection with
experimentally accessible parameters. Meanwhile, the side fringes
which normally appears in Ramsey spectroscopies are surpressed and
the dispersive shift can be minimized to negligible magnitude under
optimization with extreme ratios of dispersive duration to resonant
duration. In Sec.~\ref{sec:triple_res}, we further improve the FWHM
by 37\% by expanding into a triple-resonance scheme. Though fringes
appear in this scheme, the dispersive shift is still negligible with
suitable optimization. Conclusions are given in Sec.~\ref{sec:conclusions}.

\section{Ramsey biasing a transmon qubit\label{sec:model}}

\begin{figure}
\includegraphics[bb=0bp 0bp 360bp 327bp,clip,width=8.5cm]{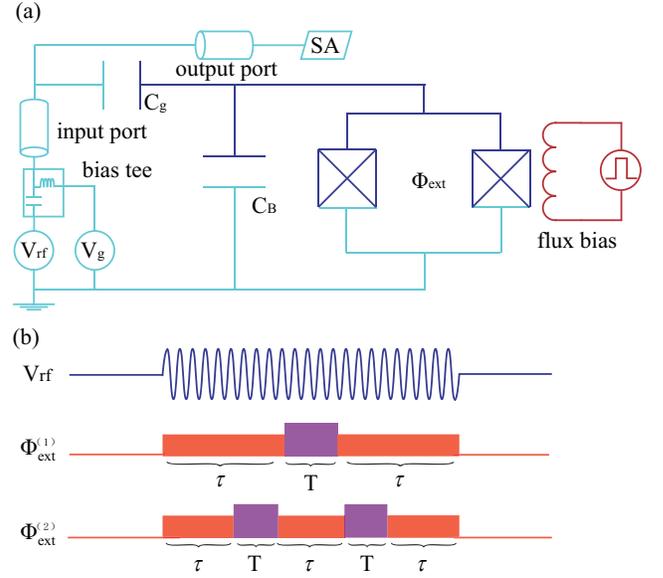}\caption{\label{fig:schematic}(a) The equivalent circuit network of the transmon
qubit and its surroundings. Two parallel Josephson junctions of junction
energy $E_{J}$ and capacitance $C_{J}$ are connected to a gate capacitor
of capacitance $C_{g}$ and shunted to the ground through a capacitor
of capacitance $C_{B}$. The transmon is biased through both a DC
offset $V_{g}$ and a magnetic flux $\Phi_{\mathrm{ext}}$. Ramsey
biasing is realized through current pulses of particular durations
in the flux bias loop. (b) The pulse sequences at the waveguide input
and at the flux bias to furnish the Ramsey resonance scheme: $V_{\mathrm{rf}}$
is set to feed a square pulse whose length matches the total time
duration of the biasing train; $\Phi_{\mathrm{ext}}^{(1)}$ illustrates
the scenario for a double-resonance scheme and $\Phi_{\mathrm{ext}}^{(2)}$
for a triple-resonance scheme. The effectiveness of the two differing
schemes for spectroscopy is depicted in the figures below.}
\end{figure}

The transmon qubit, shown in Fig.~\ref{fig:schematic}(a), is essentially
a superconducting loop containing two parallel Josephson junctions,
where one side of the loop being isolated by the gate capacitor of
capacitance $C_{g}$ and by a comb capacitor of capacitance $C_{B}$
forms a Cooper-pair box (CPB)~\cite{Koch2007}. The CPB is biased
through the charge-pair number $n_{g}$ on $C_{g}$ and through an
external magnetic flux $\Phi_{\mathrm{ext}}$ threading the loop~\cite{Schreier2008}.
Defining the charge energy $E_{C}=e^{2}/2(C_{J}+C_{B}+C_{g})$ and
the junction energy $E_{J}=I_{0}\Phi_{0}/2\pi$, the qubit Hamiltonian
reads ($\hbar=1$)
\begin{equation}
H_{\mathrm{q}}=4E_{C}\left(n-n_{g}\right)^{2}-E_{J}\left|\cos\left(\pi\frac{\Phi_{\mathrm{ext}}}{\Phi_{0}}\right)\right|\cos\varphi,\label{eq:Ham}
\end{equation}
where the canonical variables $n$ and $\varphi$ denote the electron-pair
number in the CPB and the phase difference of the tunneling current
across the junction, respectively. $\Phi_{0}$ denotes the flux quantum
$h/2e$. Operating usually at a large energy ratio of $E_{J}/E_{C}$
on the order of tens or hundreds, the transmon is optimized at the
biasing point $n_{g}=1/2$. For $n=-id/d\varphi$, at $n_{g}=1/2$,
Eq.~(\ref{eq:Ham}) is a second-order differential operator of $\varphi$,
which is diagonlizable under the bases of Mathieu functions. When
the energy ratio $E_{J}/E_{C}$ reaches a large limit, the anharmonic
system is approximately two-level under the basis set $\{\left|e\right\rangle ,\left|g\right\rangle \}$
(i.e. Mathieu functions corresponding to lowest two energies) and
its effective Hamiltonian becomes $H_{\mathrm{q}}^{\mathrm{eff}}=(\omega_{eg}/2)\sigma_{z}$
where the transition frequency is defined as~\cite{Fink2009}
\begin{equation}
\text{\ensuremath{\omega_{eg}}}(\phi)=\sqrt{8E_{C}E_{J}\left|\cos(\pi\phi)\right|}-E_{C}.\label{eq:Eigenenery}
\end{equation}
Thus, when interacting with an external pulse, the resonance frequency
of the qubit relies on the reduced magnetic flux $\phi=\Phi_{\mathrm{ext}}/\Phi_{0}$,
which can be implemented by a current pulse generator in a neighboring
flux bias loop as shown in Fig.~\ref{fig:schematic}(a).

To realize the Ramsey biasing scheme, we generate a current pulse
train in the bias loop such that $\omega_{eg}(\phi)$ of the qubit
is alternatively tuned to its resonance and dispersive regimes with
respect to an incident microwave pulse transmitting in the waveguide.
The synchronized operation between the pulse-qubit interaction and
the bias-tuning follows the illustration in Fig.~\ref{fig:schematic}(b).
With the contribution from the qubit-field interaction, the full system
Hamiltonian containing the qubit-field interaction reads

\begin{equation}
H=\Delta\sigma_{z}+\eta(\sigma_{+}+\sigma_{-})\label{eq:sys Ham}
\end{equation}
under the rotating frame $e^{i\omega t}$ of the incident probe field,
where $\Delta=(\omega_{eg}-\omega)/2$ is the detuning between the
qubit transition frequency $\omega_{eg}$ and the probe field frequency
$\omega$. The effective coupling strength $\eta$ depends on the
power of the incident microwave pulse. Eq.~(\ref{eq:sys Ham}) can
be further diagonalized to $H'=\lambda\sigma_{z}$ where $\lambda=\sqrt{\Delta^{2}+\eta^{2}}$
with transformation matrix
\begin{equation}
U=\left[\begin{array}{cc}
\cos\theta/2 & -\sin\theta/2\\
\sin\theta/2 & \cos\theta/2
\end{array}\right]
\end{equation}
and transformation angle $\theta=\tan^{-1}(\eta/\Delta)$.

Following the Hamiltonian in the transformed basis, the coefficients
$\mathbf{C}^{\prime}(t)=[C_{e}^{\prime}(t)\;C_{g}^{\prime}(t)]$ at
the initial state becomes $\mathbf{C}^{\prime}(\tau+t_{0})=\exp\{-i\lambda\sigma_{z}\tau\}\mathbf{C}^{\prime}(t_{0})$
after a time duration $\tau$. Transforming back to the bare-state
basis in the laboratory, i.e. $\mathbf{C}(\tau+t_{0})=\exp\{i\omega(\tau+t_{0})\sigma_{z}/2\}U\exp\{-i\lambda\tau\sigma_{z}\}U^{\dagger}\mathbf{C}(t_{0})$,
the state coefficients are written as
\begin{align}
C_{e}(\tau+t_{0})= & \left(\cos\lambda\tau-i\cos\theta\sin\lambda\tau\right)e^{-i\omega\tau/2}C_{e}(t_{0}).\nonumber \\
 & -i\sin\theta\sin\lambda\tau\,e^{-i\omega(\tau/2+t_{0})}C_{g}(t_{0}),\label{eq:coe_e}\\
C_{g}(\tau+t_{0})= & \left(\cos\lambda\tau+i\cos\theta\sin\lambda\tau\right)e^{i\omega\tau/2}C_{g}(t_{0})\nonumber \\
 & -i\sin\theta\sin\lambda\tau\,e^{i\omega(\tau/2+t_{0})}C_{e}(t_{0}).\label{eq:coe_g}
\end{align}

With Eqs.~(\ref{eq:coe_e})-(\ref{eq:coe_g}), we can consider the
variations of coefficients during a time duration $\tau$ when the
microwave pulse is controlling the qubit to operate in different regimes.
Under the Ramsey scheme, the first operating regime is that of resonance.
Hence, without loss of generality, we can assume the initial moment
to be $t_{0}=0$, at which the qubit adopts ground state, i.e. $C_{g}(0)=1$
and $C_{e}(0)=0$. After being biased to resonance (flux bias $\phi=0.46$)
for a duration $\tau$, the qubit has
\begin{align}
C_{e}(\tau) & =-i\sin\theta\sin\lambda\tau\,e^{-i\omega\tau/2},\label{eq:Ce_tau}\\
C_{g}(\tau) & =\left(\cos\lambda\tau+i\cos\theta\sin\lambda\tau\right)e^{i\omega\tau/2}.\label{eq:Cg_tau}
\end{align}

The second operating regime being dispersive, we consider the approximate
Hamiltonian when the qubit is far-detuned from the incident microwave,
i.e. $\Delta\gg\eta$ and thus Eq.~(\ref{eq:sys Ham}) becomes
\begin{equation}
H\approx H_{D}=\left(\Delta_{D}+\frac{\omega}{2}\right)\sigma_{z}\label{eq:dispersive Ham}
\end{equation}
where $\Delta_{D}=(\omega_{eg}^{\prime}-\omega)/2+\eta^{2}/(\omega_{eg}^{\prime}-\omega)$
is the dispersive qubit-probe detuning at flux bias $\phi'=0.49$.
The dispersive Hamiltonian only imposes a dynamic phase on the state
coefficients during evolution. At the end of a duration $T$ in the
dispersive regime after the previous operation at resonance, we then
have
\begin{equation}
C_{e,g}(\tau+T)=e^{\mp i(\Delta_{D}+\omega/2)T}C_{e,g}(\tau).\label{eq:Ceg_T}
\end{equation}

The double-resonance scheme is furnished by biasing the qubit to resonance
again for a duration $\tau$, as shown in Fig.~\ref{fig:schematic}(b).
By compounding Eqs.~(\ref{eq:Ce_tau}), (\ref{eq:Cg_tau}), and (\ref{eq:Ceg_T})
and substituting into (\ref{eq:coe_e})-(\ref{eq:coe_g}), we obtain
the coefficient of the excited state to be
\begin{multline}
C_{e}(2\tau+T)=-2ie^{-i\omega(\tau+T/2)}\sin\theta\sin\lambda\tau\\
\biggl\{\cos\lambda\tau\cos\Delta_{D}T-\cos\theta\sin\lambda\tau\sin\Delta_{D}T\biggr\}\label{eq:Ce_full}
\end{multline}
at the end of the full bias train and the transition probability across
the probe frequency $\omega$ is then computable as $P_{e}=|C_{e}|^{2}$.

\section{Spectroscopy under double-resonance\label{sec:double_res}}

Following the expression of Eq.~(\ref{eq:Ce_full}), $P_{e}$ is
a function of the time lengths $\tau$ and $T$ in addition to the
probe frequency $\omega$. For a double-resonance scheme, $2\tau+T$
would be the total probe pulse length and the measurement reading
would be the attenuated pulse power of the same length received by
the spectrum analyzer (Cf. Fig.~\ref{fig:schematic}). Therefore,
viewing them as free variables, we can optimize over $\tau$ and $T$
and remove them from the equation. The optimization goal is maximal
$P_{e}$ at the resonance frequency with minimum FWHM for the linewidth.

To implement the optimization, we consider $T=R\tau$, i.e. $R$ is
a proportional constant indicating the dispersion to resonance time
ratio, and that the sampling duration parametrized by $\tau$ is stochastically
distributed, which gives the observed transition probability as the
time integral of $P_{e}$. The distribution typically adopted for
spectroscopic purposes is that of Maxwell~\cite{Ramsey1950,Walstad2013},
\begin{equation}
p(x)=e^{-x^{2}}x^{3}.\label{eq:maxwell distri}
\end{equation}
To make the distribution compatible with the expression of $P_{eg}$,
we let the dimensionless $x=\tau/s$, where $s$ is a time constant
to be determined. Further, we write $|C_{e}|^{2}$ as a function of
$\tau$ in terms of a sum of cosines to simplify the integrand. This
is possible when we observe from Eq.~(\ref{eq:Ce_full}) that the
factors containing $\lambda\tau$, $\omega\tau$, and $R\Delta_{D}\tau$
can all be reduced to cosines or combinations of cosines and constants.

Therefore, in terms of the integral

\begin{equation}
I_{s}(\beta)=\int_{0}^{\infty}dx\,e^{-x^{2}}x^{3}\cos(2\beta sx),\label{eq:int_cosine}
\end{equation}
where $sx$ is equivalent to the time variable and $\beta$ denotes
the various frequency terms, the weighted average of $|C_{e}|^{2}$
is 
\begin{align}
 & \left\langle P_{e}\right\rangle =\frac{F'^{4}+4F^{2}}{4}+\frac{F'^{4}-2F^{2}}{2}I_{s}(R\Delta_{D})\nonumber \\
 & -2F^{2}I_{s}(\lambda)-\frac{F'^{4}}{2}I_{s}(2\lambda)\nonumber \\
 & +\left(F+F'\right)\left[FI_{s}\left(\lambda+R\Delta_{D}\right)-\frac{(F+F')}{4}I_{s}\left(2\lambda+R\Delta_{D}\right)\right]\nonumber \\
 & +\left(F-F'\right)\left[FI_{s}\left(\lambda-R\Delta_{D}\right)-\frac{(F-F')}{4}I_{s}\left(2\lambda-R\Delta_{D}\right)\right]\label{eq:Pe_avg}
\end{align}
where we used the abbreviations
\begin{align}
F & =\frac{\eta\Delta}{\lambda^{2}},\qquad F'=\frac{\eta}{\lambda}.
\end{align}
To optimize for a maximal $\left\langle P_{e}(\omega)\right\rangle $
at the qubit resonance frequency, we consider tuning the integral
$I_{s}$, which contribute both positive and negative terms in Eq.~(\ref{eq:Pe_avg}).
Since the constant term in Eq.~(\ref{eq:Pe_avg}) is positive, we
minimize the magnitude of $I_{s}$ to opt for a maximal $\left\langle P_{e}\right\rangle $
overall. Following the expression of Eq.~(\ref{eq:int_cosine}),
the integrand of $I_{s}$ obtains its minimum value at $\beta s\approx0.68\pi$.
Further, about close-resonance limit where we regard $\Delta=0$ and
thus $\lambda=\eta$, Eq.~(\ref{eq:Pe_avg}) contains the integrals
$I_{s}(\eta)$ and $I_{s}(2\eta)$, whose coefficients are negative,
in addition to the $\omega$-dependent integrals. The optimization
is obtained, therefore, by minimizing these two $\omega$-independent
integrals. Obviously, they cannot be simultaneously minimized at the
same $s$ value, so we determine numerically the compromised time
constant $s$ that leads to a maximal peak value with minimal FWHM.

\begin{figure}
\includegraphics[bb=10bp 0bp 480bp 366bp,clip,width=8.5cm]{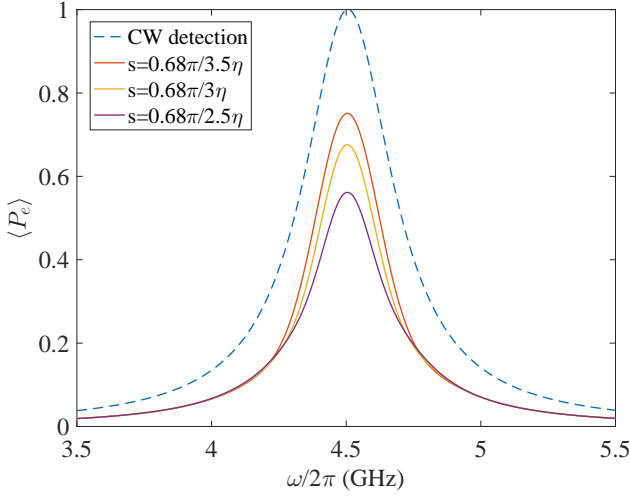}

\caption{(Color online) The transition probability of a transmon qubit as a
function of probe field frequency $\omega/2\pi$ at three values of
$s$ between $0.68\pi/3.5\eta$ and $0.68\pi/2.5\eta$, given as the
family of solid curves. The blue dashed line, which corresponds to
the spectrum of CW detection under the same system parameters, serves
as a reference.~\label{fig:probability 2tau+T}}
\end{figure}

Figure~\ref{fig:probability 2tau+T} plots the transition probability
$\left\langle P_{e}\right\rangle $ of a transmon qubit against the
probe frequency $\omega$ at various values of $s$ under the proposed
Ramsey-biased scheme. The energy ratio $E_{J}/E_{C}$ of the transmon
is set to $100$. The two distinct flux biases during the double-resonance
process are given as above. The coupling strength is assumed to be
$\eta/2\pi=100\mathrm{MHz}$. The resonance peak, which occurs at
$4.505\mathrm{GHz}$, is obtained at the optimized $T$ to $\tau$
ratio $R=0.001$. Compared to the transition spectrum according to
CW detection (shown as the dashed curve), the new scheme provides
an improved FWHM at all values of $s$. In particular, at $s=0.68\pi/3\eta$,
FWHM is reduced to $306\mathrm{MHz}$ from $400\mathrm{MHz}$ of the
CW detection ($\approx23\%$ reduction), given the cost of a slightly
reduced peak magnitude, demonstrating the advantage of the Ramsey-biased
scheme. For all values of $s$, the shift in frequency due to dispersion
is about $1.7$~MHz from CW detection, which is almost negligible
compared to the resonance frequency.

\section{Spectroscopy under triple-resonance\label{sec:triple_res}}

To further optimize the transition spectrum, i.e. to obtain a curve
with a narrower linewidth to those shown in Fig.~\ref{fig:probability 2tau+T},
we consider a triple-resonance scheme. Instead of the $\tau$-$T$-$\tau$
segmentation for the qubit-pulse interaction, we consider three segments
of $\tau$, during which the qubit is resonant, and two segments of
$T$, during which the qubit is dispersive. Again, the two operating
regimes are interlacing with each other and we can reuse Eqs.~(\ref{eq:coe_e})-(\ref{eq:coe_g})
and (\ref{eq:Ceg_T}). Applying them on Eqs.~(\ref{eq:Ce_full}),
we obtain the coefficient 
\begin{align}
C_{e}(3\tau+2T)= & -ie^{-i\omega(3\tau/2+T)}\sin\theta\sin\lambda\tau\times\nonumber \\
 & \Bigl\{2\left(\cos^{2}\lambda\tau-\cos^{2}\theta\sin^{2}\lambda\tau\right)\cos2\Delta_{D}T\nonumber \\
 & -2\cos\theta\sin2\lambda\tau\sin2\Delta_{D}T\nonumber \\
 & +\cos^{2}\lambda\tau+\cos2\theta\sin^{2}\lambda\tau\Bigr\}.\label{eq:Ce_triple}
\end{align}
of the excited state at the end of the pulse train.

\begin{figure}
\includegraphics[bb=15bp 0bp 470bp 483bp,clip,width=8.5cm]{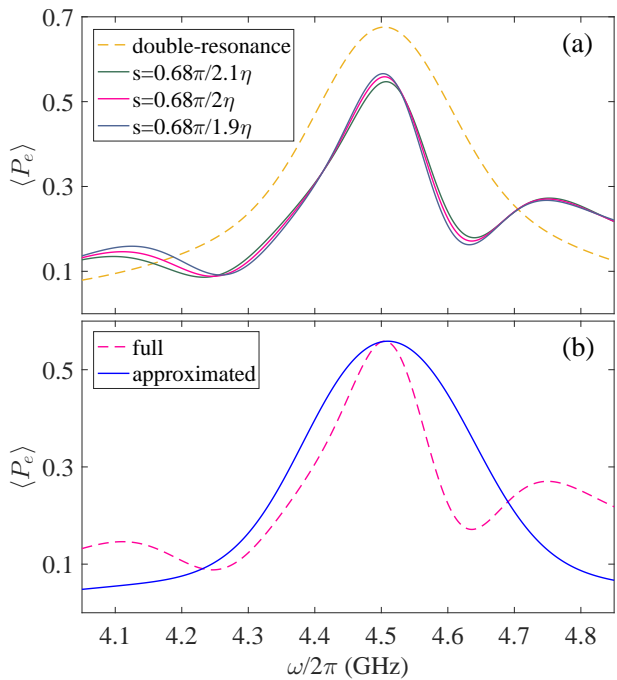}

\caption{(Color online) (a) The transition probability under the triple-resonance
scheme at three different values of $s$ (solid curves) compared to
the double-resonance scheme (dashed curve, $s=0.68\pi/3\eta$). (b)
$\left\langle P_{e}^{\mathrm{res}}\right\rangle $ under close-resonance
approximation $s=0.68\pi/2\eta$ (solid curve) compared to one without
approximation (dashed curve).~\label{fig:triple_res}}
\end{figure}

The expression of $C_{e}$ is much more complicated for the triple
resonance scheme than that given in Eq.~(\ref{eq:Ce_full}). Before
reducing the terms of $|C_{e}|^{2}$ to cosines to carry out the integration
in Eq.~(\ref{eq:int_cosine}), we simplify the expression of Eq.~(\ref{eq:Ce_triple})
to examine the analytical dependence only in the close-resonance region.
We again let $T=R\tau$ ($R$ to be determined similarly to Sec.~\ref{sec:double_res})
and regard $C_{e}$ again as a function of $\tau$, which will be
integrated out. At $\Delta\to0$ such that all $\cos^{n}\theta$ terms
vanish, we have
\begin{align}
\left\langle P_{e}^{\mathrm{res}}\right\rangle  & =\frac{F'^{2}}{16}\Bigl\{6-10I_{s}(\lambda)+4I_{s}(2\lambda)-6I_{s}(3\lambda)\nonumber \\
 & +4I_{s}(2R\Delta_{D})+4I_{s}(\lambda+R\Delta_{D})+4I_{s}(\lambda-R\Delta_{D})\nonumber \\
 & +I_{s}(\lambda+2R\Delta_{D})+I_{s}(\lambda-2R\Delta_{D})\nonumber \\
 & -2I_{s}(2\lambda+2R\Delta_{D})-2I_{s}(2\lambda-2R\Delta_{D})\nonumber \\
 & -4I_{s}(3\lambda+R\Delta_{D})-4I_{s}(3\lambda-R\Delta_{D})\nonumber \\
 & -I_{s}(3\lambda+2R\Delta_{D})-I_{s}(3\lambda-2R\Delta_{D})\Bigr\}.\label{eq:Pe_avg_triple}
\end{align}
Comparing Eq.~(\ref{eq:Pe_avg_triple}) with Eq.~(\ref{eq:Pe_avg}),
one sees that the extra segment of resonance effectively contributes
extra terms of frequency $3\lambda$. Like $2\lambda$-frequency terms
in Eq.~(\ref{eq:Pe_avg}), these triple-frequency terms all have
negative weights, thus reducing the probability magnitude at frequencies
distant from resonance. Cutting down the magnitude naturally leads
to a narrower resonance peak. $\left\langle P_{e}\right\rangle $
computed directly as $|C_{e}|^{2}$ under the triple-resonance scheme
is plotted in Fig.~\ref{fig:triple_res}(a) for three optimized values
of $s$, against the double-resonance result for $s=0.68\pi/3\eta$.
The optimization is obtained under $T$ to $\tau$ ratio $R=0.045$.
For $s=0.68\pi/2\eta$, the triple-resonance scheme has a FWHM of
$193$~MHz, i.e. a further $37\%$ reduction over the double-resonance
case, albeit the compromise at the peak absorption magnitude.

The resonance peak, like that of Fig.~\ref{fig:probability 2tau+T},
experiences negligible dispersion shift but exhibits sideband fringes.
These fringes, unlike those in the original Ramsey spectroscopy scheme,
are flattened and asymmetric. To examine the effectiveness of $\left\langle P_{e}^{\mathrm{res}}\right\rangle $,
it is plotted against the full numerical result in Fig.~\ref{fig:triple_res}(b)
with the same values of $s$ and $R$. The approximation coincides
with the analytical expression at close-resonance range but falls
short at estimating the spectral width. At off-resonance range, the
terms associated with high-order $\cos^{n}\theta$ coefficients are
integrals $I_{s}$ of frequency higher than $3\lambda$. Since these
terms are also negative, $\left\langle P_{e}\right\rangle $ at frequencies
distant from resonance falls off more sharply than that expected by
Eq.~\ref{eq:Pe_avg_triple}, which omit these terms. Consequently,
the fringes that are caused by higher-order secondary photon processes
are also omitted.

\section{Conclusions\label{sec:conclusions}}

We generalize Ramsey's spectroscopic method of transition detection,
which was applied to atoms traveling through optical cavities, to
detecting superconducting qubits on solid-state circuits. In this
multi-segment resonance scheme, the time segment originally reserved
for free atomic evolution is replaced by a dispersion operation for
the qubit, which give rises not only to a narrowed linewidth of transition
spectrum, but also to an elimination of side fringes. We have proved
that the unwanted frequency shift in the detection due to dispersion
is negligible when the ratio of time segment arrangments are suitably
optimized. The linewidth narrowing can be further improved when extra
resonance segments are added.
\begin{acknowledgments}
H. I. acknowledges the support by FDCT of Macau under grant 065/2016/A2
and University of Macau under grant MYRG2018-00088-IAPME. 
\end{acknowledgments}


\begin{thebibliography}{10}
\bibitem{Mooij1999}J. E. Mooij, T. P. Orlando, L. Levitov, L. Tian,
C. H. van der Wal, and S. Lloyd, Science \textbf{285}, 1036 (1999).

\bibitem{Clarke2008}J. Clarke and F. K. Wilhelm, Nature \textbf{453},
1031 (2008).

\bibitem{nakamura97}Y. Nakamura, C. D. Chen, and J. S. Tsai, Phys.
Rev. Lett. \textbf{79}, 2328 (1997).

\bibitem{Orlando1999}T. P. Orlando,\emph{ et al}., Phys. Rev. B \textbf{60},
15398 (1999).

\bibitem{vanderWal00}C. H. van der Wal, \emph{et al.} , Science \textbf{290},
773 (2000).

\bibitem{martinis02}J. M. Martinis, S. Nam, J. Aumentado, and C.
Urbina, Phys. Rev. Lett. \textbf{89}, 117901 (2002).

\bibitem{Kastner1993}M. A. Kastner, Phys. Today \textbf{46}, 24 (1993).

\bibitem{you05}J. Q. You and F. Nori, Phys. Today \textbf{58}, 42
(2005).

\bibitem{Chiorescu2003}I. Chiorescu, Y. Nakamura, C. J. P. M. Harmans,
and J. E. Mooij, Science \textbf{299}, 1869 (2003).

\bibitem{Wallraff2004}A. Wallraff, \emph{et al}., Nature \textbf{431},
162 (2004).

\bibitem{shen05}J.-T. Shen and S. Fan, Phys. Rev. Lett. \textbf{95},
213001 (2005).

\bibitem{Hime2006}T. Hime, \emph{et al}., Science \textbf{314}, 1427
(2006).

\bibitem{Shevchenko2008}S. N. Shevchenko, \emph{et al.}, Phys. Rev.
B \textbf{78}, 174527 (2008).

\bibitem{Torrey1941}H. C. Torrey, Phys. Rev. \textbf{59}, 293 (1941).

\bibitem{Ramsey1950}N. F. Ramsey, Phys. Rev. \textbf{78}, 695 (1950).

\bibitem{Abele2010}H. Abele, T. Jenke, H. Leeb, and J. Schmiedmayer,
Phys. Rev. D \textbf{81}, 065019 (2010).

\bibitem{Zanon2015}T. Zanon-Willette, V. I. Yudin, and A. V. Taichenachev,
Phys. Rev. A \textbf{92}, 023416 (2015).

\bibitem{Koch2007}J. Koch, \emph{et al}., Phys. Rev. A \textbf{76},
042319 (2007).

\bibitem{Filipp2009}S. Filipp, \emph{et al.}, Phys. Rev. Lett. \textbf{102},
200402 (2009).

\bibitem{Walstad2013}A. Walstad, American Journal of Physics \textbf{81},
555 (2013).

\bibitem{fyan12}F. Yan, \emph{et al.}, Phys. Rev. B \textbf{85},
174521 (2012).

\bibitem{fyan13}F. Yan, \emph{et al.}, Nat. Commun. \textbf{4}, 1
(2013). 

\bibitem{Schreier2008}J. A. Schreier, \emph{et al}., Phys. Rev. B
\textbf{77}, 180502 (2008).

\bibitem{Fink2009}J. M. Fink, \emph{et al}., Phys. Rev. Lett. \textbf{103},
083601 (2009).
\end{thebibliography}
\end{document}